\documentclass[aps,twocolumn,showpacs]{revtex4}
\usepackage{amsmath}
\usepackage{graphicx}

\begin{document}

\title{Classical Analytical Mechanics and Entropy Production}
\author{J. Silverberg and A. Widom}
\affiliation{Physics Department, Northeastern University, Boston MA 02115}

\begin{abstract}
The usual canonical Hamiltonian or Lagrangian formalism of classical mechanics 
applied to macroscopic systems describes energy conserving adiabatic motion.
If irreversible diabatic processes are to be included, then the law of increasing 
entropy must also be considered. The notion of entropy then enters into the general 
classical mechanical formalism. The resulting general formulation and its 
physical consequences are explored.  
\end{abstract}
\maketitle

\section{\label{intro} Introduction}

In typical macroscopic classical mechanics problems, one deals with 
Hamiltonians describing a few degrees of freedom, say 
\begin{math} (P,Q)=(P_1,\ldots, P_n,Q_1,\ldots, Q_n) \end{math}. From 
a microscopic viewpoint, the Hamiltonian contains an enormous number 
of degrees of freedom, say \begin{math} N\sim 10^{23}  \end{math}. To 
derive the macroscopic Hamiltonian from the microscopic Hamiltonian 
requires the notion of thermodynamic entropy. In detail, suppose a microscopic 
Hamiltonian \begin{math} {\cal H}(P,Q,p,q)  \end{math} wherein 
\begin{math} (p,q) \end{math} represent microscopic degrees of freedom. 
Discrete degrees of freedom are left implicit. For fixed classical macroscopic 
values for the reduced degrees of freedom \begin{math} (P,Q) \end{math}, one 
may consider the quantum microscopic energy eigenvalue problem \cite{LandauQM:2001}
\begin{eqnarray}
{\cal H}\left(P,Q,p=-i\hbar \frac{\partial}{\partial q},q\right)\psi_k(q)
=E \psi_k(q) ,
\nonumber \\ 
k=1,2,\ldots,\Omega .
\label{intro1}
\end{eqnarray}
The thermodynamic entropy is determined by the microscopic degeneracy via 
the Boltzmann-Gibbs law \cite{LandauSP:2001}
\begin{equation}
S(E,P,Q)=k_B\ln \Omega (E,P,Q).
\label{intro2}
\end{equation}
In principle, Eq.(\ref{intro2}) may be solved for the energy in the form 
\begin{equation}
E=H(P,Q,S),
\label{intro3}
\end{equation}
yielding the classical macroscopic Hamiltonian which is now an explicit 
function of entropy. Our purpose is to consider the physical consequences 
of including the entropy in classical canonical dynamics. 

In Sec.\ref{Sconst}, we consider the adiabatic classical dynamics of 
macroscopic systems. Since the entropy \begin{math} S \end{math} is uniform 
in time for adiabatic dynamics, the usual Hamiltonian and Lagrangian dynamics 
holds true, albeit with a macroscopic Lagrangian 
\begin{math} L(\dot{Q},Q,S)  \end{math}
which also explicitly depends on entropy. In adiabatic classical mechanics, 
both entropy and energy are conserved. 

In Sec.\ref{Sincrease}, we explore the consequences of diabatic processes 
wherein the entropy is increasing. In accordance with the first and second 
laws of thermodynamics the system energy is still conserved albeit some of 
the energy is converted into heat precisely defined by the increase in entropy.
In detail, the Hamiltonian in Eq.(\ref{intro3}) along with friction force 
components \begin{math} (f_1,\ldots ,f_n)  \end{math} determine the coupled 
equations of motion
\begin{eqnarray}
\dot{Q}^a=\frac{\partial H(P,Q,S)}{\partial P_a}\ ,
\nonumber \\ 
\dot{P}_a=-\frac{\partial H(P,Q,S)}{\partial Q^a}+f_a(P,Q,S)\ ,
\nonumber \\ 
T\dot{S}=-\dot{Q}^af_a \ge 0.
\label{intro4}
\end{eqnarray}
To assure an increasing entropy, the friction force components must be opposite 
to the velocity components. While the heating rate due to frictional forces 
\begin{math} T\dot{S}  \end{math} is positive, energy is nevertheless strictly 
conserved, i.e. 
\begin{equation} 
\dot{E} \equiv \dot{H}=0
\label{intro5}
\end{equation}
in virtue of Eqs.(\ref{intro4}).

In Sec.\ref{GenFric}, well known cases of friction 
are discussed.  The friction term is then generalized
to a function of coordinates and velocities as 
\begin{math} f_a=-R(Q,\dot{Q})v^a \end{math},
and the consequences are explored. 

In Sec.\ref{UO} the example of the 
underdamped simple harmonic oscillator 
is considered. It is used as an pedagalogical demonstration in the 
calculation purely thermodunamic quantities such as temperature as fucntions 
of time for the formalism considered herein.

\section{Adiabatic Processes \label{Sconst}}

In an adiabatic process, the system entropy is uniform in time
\begin{equation}
\dot{S}=0.
\label{Sconst1}
\end{equation}
In general, the first and second laws of thermodynamics may be written 
in the form \cite{Fermi:1956}
\begin{eqnarray}
dH=V^a dP_a-F_adQ^a+TdS, 
\nonumber \\ 
V^a (P.Q,S)=\frac{\partial H(P,Q,S)}{\partial P_a}\ ,
\nonumber \\ 
F_a (P.Q,S)=-\frac{\partial H(P,Q,S)}{\partial Q^a}\ ,
\nonumber \\ 
T(P,Q,S)=\frac{\partial H(P,Q,S)}{\partial S}\ ,
\label{Sconst2}
\end{eqnarray}
wherein the Einstein summation convention over index {\it a} is being employed, and 
\begin{math} T \end{math} is the system temperature.
Adiabatic macroscopic Hamiltonian dynamics is of the usual classical form
\begin{equation}
\dot{Q}^a=V^a (P.Q,S),\ \ 
\dot{P}_a=F_a (P.Q,S)\ \ {\rm and} 
\ \ \dot{S}=0.
\label{Sconst3}
\end{equation}

To obtain the equivalent Lagrangian form of the equations of motion, 
note that 
\begin{equation}
L=-H+V^aP_a.
\label{Sconst4}
\end{equation}
Eqs.(\ref{Sconst2}), (\ref{Sconst3}) and (\ref{Sconst4}) imply 
\begin{equation}
dL=P_a d\dot{Q}^a+F_a dQ^a-TdS.
\label{Sconst5}
\end{equation}
The adiabatic equations of motion in Lagrangian form are 
\begin{equation}
\frac{d}{dt}\left(\frac{\partial L}{\partial \dot{Q}^a}\right)_{Q,S}
=\left(\frac{\partial L}{\partial Q^a}\right)_{\dot{Q},S},
\ \ {\rm and}\ \ \dot{S}=0.
\label{Sconst6}
\end{equation}
The system temperature in the Lagrangian representation 
follows from Eq.(\ref{Sconst5}) according to 
\begin{equation}
T=-\left(\frac{\partial L}{\partial S}\right)_{\dot{Q},Q}\ .
\label{Sconst7}
\end{equation}
Conservation of energy,
\begin{eqnarray}
E=\dot{Q}^a\left(\frac{\partial L}{\partial \dot{Q}^a}\right)_{Q,S}-L
\nonumber \\ 
\dot{E}=0,
\label{Sconst8}
\end{eqnarray}
follows directly from Eqs.(\ref{Sconst6}).

Both energy and entropy are conserved during a classical adiabatic motion. 
When forces of friction are present, the energy is still conserved but the 
entropy increases with time in agreement with the first and second laws of 
thermodynamics. Let us see how this comes about.

\section{Diabatic Processes \label{Sincrease}}

Our purpose is to derive the rule of conservation of energy in the presence 
of frictional forces. Starting from the Hamiltonian definition of energy  
\begin{math} E=H(P,Q,S) \end{math}, one may compute the time 
derivative 
\begin{equation}
\dot{E}=\frac{\partial H}{\partial P_a} \dot{P}_a 
+ \frac{\partial H}{\partial Q^a} \dot{Q}^a
 + \frac{\partial H}{\partial S} \dot{S}.
\label{Sincrease1}
\end{equation}
Employing Eq.(\ref{intro4}) for frictional forces yields  
\begin{equation}
\dot{E} =  \dot{Q}^a\dot{P}_a + (-\dot{P}_a + f_a)\dot{Q}^a + T\dot{S} 
= f_a \dot{Q}^a + T\dot{S}.
\label{Sincrease2}
\end{equation}
Energy conservation in the form  
\begin{equation}
\dot{E}=0,
\label{Sincrease3}
\end{equation}
is equivalent to the condition that frictional forces 
produce heat according to the entropy rule 
\begin{equation}
T \dot{S}=-f_a \dot{Q}^a .
\label{Sincrease4}
\end{equation}
The components of the frictional force 
\begin{math} f_a \ \ (a=1 \ldots n) \end{math} will oppose the direction of motion
\begin{math} \dot{Q}^a \ \ (a=1 \ldots n) \end{math} giving rise to the second law 
of thermodynamics in the form 
\begin{equation}
\dot{S}\ge 0.
\label{Sincrease5}
\end{equation}

The  equations of motion in Lagrangian form with diabatic processes can be 
found by extending Eq.(\ref{Sconst6}) to the case wherein 
\begin{math} \dot{S} \ne 0 \end{math}.  This may be done by directly adding the 
frictional forces to Lagrange's equations, which now read
\begin{eqnarray}
\frac{d}{dt}\left(\frac{\partial L}{\partial \dot{Q}^a}\right) = 
 \frac{\partial L}{\partial Q^a} + f_a,
\nonumber \\
P_a(Q^a,\dot{Q}^a,S) = \frac{\partial L(Q^a,\dot{Q}^a,S)}{\partial \dot{Q}^a} ,
\nonumber \\
F_a(Q^a,\dot{Q}^a,S) = \frac{\partial L(Q^a,\dot{Q}^a,S)}{\partial Q^a}.
\label{Sincrease6}
\end{eqnarray}
More compactly, Eq(\ref{Sincrease6}) asserts that 
\begin{equation}
\dot{P}_a=F_a + f_a.
\label{Sincrease6a}
\end{equation}
The energy expression from the entropy dependent Lagrangian,
\begin{eqnarray}
E=E(Q,\dot{Q},S),
\nonumber \\ 
E=\dot{Q}^a \left(\frac{\partial L}{\partial \dot{Q}^a}\right)-L,
\label{Sincrease7}
\end{eqnarray}
may be shown, via Eqs.(\ref{Sconst5}) and (\ref{Sincrease6}), to be consistent with 
\begin{eqnarray}
E=H(P,Q,S)
\nonumber \\ 
dE=\dot{Q}^adP_a -F_adQ^a + TdS.
\label{Sincrease8}
\end{eqnarray}
Employing the energy conservation Eq.(\ref{Sincrease3}) and
the diabatic Lagrange's Eq.(\ref{Sincrease6a}), one again finds the heating rate 
\begin{equation}
T \dot{S}=-f_a \dot{Q}^a .
\label{Sincrease9}
\end{equation}
The above considerations show that the same rule for entropy production follows from 
{\em both} the Lagrangian and Hamiltonian formalisms.

\section{The General Friction Function \label{GenFric}}

We now examine in more detail the generalized frictional force 
\begin{math} f_a \end{math}.  In theory and experiment, it is known that friction 
is generally a complicated phenomenon requiring regime-specific 
models and free phenomenology parameters \cite{McMillan:1997} \cite{Drexel:2001}.
For example, consider the 
differences between dry friction, viscous drag and atmospheric re-entry.  
Respectively, the dissipative force is constant, linear and quadratic 
in velocity.  Given this assortment of possibilities, a general approach must 
have enough flexibility to describe a wide range of phenomena while still 
conforming to certain physically intuitive notions.  It is 
clear that any friction force would oppose the direction of velocity.  
It should also depend on the generalized coordinates such that the expressions  
involve tensors under arbitrary coordinate transformation; e.g.  
\begin{equation}
f_a(Q,\dot{Q}) = -R_{ab}(Q,\dot{Q}) \dot{Q}^b
\label{Rayleigh1}
\end{equation}
satisfies these conditions.  

Examples of transport tensors include 
\begin{eqnarray}
R_{ab}=\frac{\alpha_{ab}(Q)}
{\sqrt{g_{df}(Q)\dot{Q}^d \dot{Q}^f}}\ \ \ {\rm (dry\ friction)}, 
\nonumber \\ 
R_{ab}=\eta_{ab}(Q)\ \ \ {\rm (viscous\ drag)},
\nonumber \\ 
R_{ab}=\frac{\beta_{abcd}(Q)\dot{Q}^c\dot{Q}^d}
{\sqrt{g_{fg}(Q)\dot{Q}^f \dot{Q}^g}}\ \ \ {\rm (re-entry)}.
\label{Rayleigh2}
\end{eqnarray}
In all of the above cases,  
the entropy production from Eq.(\ref{Sincrease4}) may be written 
\begin{equation}
T\dot{S}=R_{ab}(Q,\dot{Q})\dot{Q}^a\dot{Q}^b .
\label{Rayleigh3}
\end{equation}

In the case of viscous drag,
\begin{equation}
f_a(Q,\dot{Q})=-\eta_{ab}(Q)\dot{Q}^b
\label{Rayleigh4}
\end{equation}
which can be written as the derivative with respect to velocity of some
function $\Gamma(Q,\dot{Q})$,
\begin{equation}
f_a(Q,\dot{Q})=-\frac{\partial \Gamma(Q,\dot{Q})}{\partial \dot{Q}^a}.
\label{Rayleigh5}
\end{equation}
Hence the expression Eq.(\ref{Rayleigh5}) can be written in the quadratic form
\begin{equation}
\Gamma(Q,\dot{Q})=\frac{1}{2}\eta_{ab}(Q)\dot{Q}^a\dot{Q}^b,
\label{Rayleigh6}
\end{equation}
which is otherwise commonly known as the Rayleigh dissipation function 
\cite{Rayleigh:1945} \cite{LandauM:2001} \cite{Morabito:2004}.

\section{The Underdampened Oscillator \label{UO}}

Let us illustrate for a simple soluble model how the thermodynamic 
variables may change in time along with the purely mechanical variables. 
For this purpose we consider the underdamped simple harmonic oscillator. 
The Lagrangian is 
\begin{equation}
L=\frac{m}{2}\left(\dot{x}^2-\omega_0^2x^2\right),
\label{UO1}
\end{equation}
with an entropy production of 
\begin{equation}
T\dot{S}=2m\gamma \dot{x}^2 \ \ {\rm wherein}\ \ \ \gamma < \omega_0.
\label{UO2}
\end{equation}
The equation of motion for the underdamped oscillator,   
\begin{equation}
\ddot{x}+2\gamma \dot{x}+\omega_0^2=0,
\label{UO3}
\end{equation}
has the well known solution 
\begin{equation}
x=x_0 e^{-\gamma t}\cos(\Omega t+\phi )
\ \ \ {\rm wherein}\ \ \ \Omega=\sqrt{\omega_0^2-\gamma^2}.
\label{UO4}
\end{equation}

\begin{figure}[bp]
\scalebox {0.6}{\includegraphics{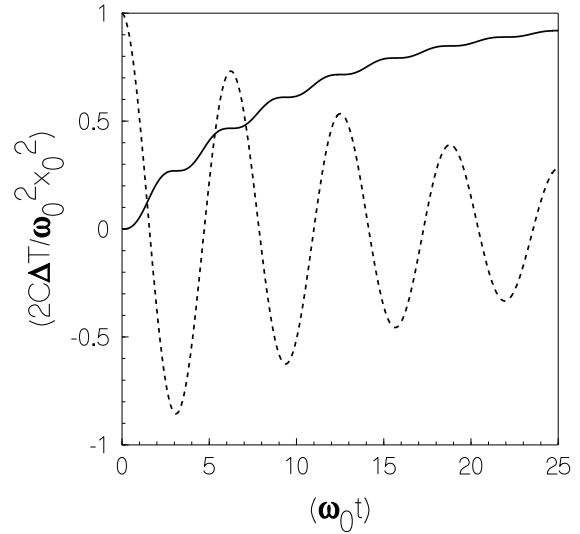}}
\caption{For an undamped oscillator with a quality factor 
$Q=\gamma/\omega_0 = 20.0$, plotted as a ``dotted'' curve 
is the relative displacement $x/x_0$ as a function of time.
Plotted as a solid curve is the temperature change 
$\Delta T/(T_\infty -T_0)$ wherein $T_\infty $ is the final 
equilibrium temperature after the oscillator stops moving.}
\label{Fig1}
\end{figure}

If the harmonic oscillator has a constant specific heat 
\begin{math} C \end{math}, 
\begin{equation}
mC=T\frac{dS}{dT}\ ,
\label{UO5}
\end{equation}
then Eqs.(\ref{UO2}), (\ref{UO4}) and (\ref{UO5}) allow one to compute 
how the temperature of the oscillator will vary in time; i.e. 
\begin{equation}
T(t)=T_0+\left(\frac{2\gamma }{C}\right)\int_0^t\dot{x}(t^\prime )^2dt^\prime .
\label{UO6}
\end{equation}
For the case of zero initial phase \begin{math} \phi=0 \end{math},
\begin{equation}
\lim_{t\to \infty} T(t)\equiv T_0+T_\infty =T_0+\frac{m}{2}\omega_0^2x_0^2,
\label{UO6a}
\end{equation}
one finds that the temperature increase \begin{math} \Delta T=T-T_0 \end{math} obeys 
\begin{eqnarray}
\frac{2C\Delta T(t)}{\omega_0^2x_0^2}=1-e^{-2\omega_0t/Q}\times 
\ \ \ \ \ \ \ \ \ \ \ 
\nonumber \\
\left[1+\frac{\cos(2\Omega t)-1}{Q^4}
+\left(1-\frac{1}{Q^2}\right)^{3/2}\frac{\sin(2\Omega t)}{Q}\right],
\label{UO7}
\end{eqnarray}
where the oscillation quality factor \begin{math} Q=(\omega_0/\gamma )\end{math}.
We have plotted both the damped harmonic oscillator coordinate and the the temperature 
increase \begin{math} \Delta T \end{math} in Fig.\ref{Fig1} above as an illustration 
of how thermodynamic parameters as functions of time can be calculated similarly to computations 
of ordinary mechanical coordinates.

\section{Conclusion \label{conc}}

The classical formalism of mechanics was extended to also include 
consideration of entropy.  The results allowed for a revised form
of Lagrange's and Hamilton's equations which necessarily included 
energy dissipation due to frictional forces.  It was shown that when the frictional
forces vanished, that the entropy production was also $0$ and classical results were 
reproduced.  However, in the diabatic cases, the law of energy conservation
gave rise to the conventional condition of entropy production in both the Lagrangian and
Hamiltonian frameworks. A simple example was then considered and worked out in detail
where the rate of entropy production was connection to the heating rate.  The 
explicit connection between the motion of a body and the rise in temperature 
was shown in closed analytical form. A more complicated problem such as the entry 
into the atmosphere and burn of a meteorite can in principle be treated by the methods 
here discussed although analytical solutions would appear unlikely.

\end{document}